\documentclass[aps,prl,twocolumn,superscriptaddress]{revtex4-1}% Á½ÐÐ
\usepackage{epsf}
\usepackage{amsfonts}
\usepackage[fleqn]{amsmath}
\usepackage{amssymb,yfonts,mathrsfs,bbm}
\usepackage{graphicx}
\usepackage{longtable}
\usepackage{amsmath}
\usepackage{graphicx }
\usepackage{graphics}
\usepackage{psfrag}
\usepackage[]{caption2} 
\usepackage{color}
\definecolor{purple}{rgb}{0.7,0.0,0.7}
\definecolor{orange}{rgb}{1,0.65,0.0}
\definecolor{dgreen}{rgb}{0.3, 0.6, 0.1}

\begin{document}

\title{The Interlayer Shear Modes in Twisted Multi-layer Graphenes: Interlayer coupling, Davydov Splitting and Intensity Resonance}
%\title{Interface Coupling and Shear Mode Intensity Resonance in Twisted Multilayer Graphene}

\author{Jiang-Bin Wu}
\author{Xin Zhang}
\affiliation{State Key Laboratory of Superlattices and Microstructures, Institute of Semiconductors, Chinese Academy of Sciences, Beijing 100083, China}
\author{Mari Ij\"{a}s}
\affiliation{Cambridge Graphene Centre, University of Cambridge, Cambridge CB3 0FA, UK}
\author{Wen-Peng Han}
\author{Xiao-Fen Qiao}
\author{Xiao-Li Li}
\author{De-Sheng Jiang}
\affiliation{State Key Laboratory of Superlattices and Microstructures, Institute of Semiconductors, Chinese Academy of Sciences, Beijing 100083, China}
\author{Andrea C. Ferrari}
\affiliation{Cambridge Graphene Centre, University of Cambridge, Cambridge CB3 0FA, UK}
\author{Ping-Heng Tan}
\affiliation{State Key Laboratory of Superlattices and Microstructures, Institute of Semiconductors, Chinese Academy of Sciences, Beijing 100083, China}

% \date{\today}

\begin{abstract}
{Graphene and other two-dimensional crystals can be combined to form various hybrids and heterostructures, creating materials on demand, in which the interlayer coupling at the interface leads to modified physical properties as compared to their constituents. Here, by measuring Raman spectra of shear modes, we probe the coupling at the interface between two artificially-stacked few-layer graphenes rotated with respect to each other. The strength of interlayer coupling between the two interface layers is found to be only 20\% of that between Bernal-stacked layers. Nevertheless, this weak coupling manifests itself in a Davydov splitting of the shear mode frequencies in systems consisting of two equivalent graphene multilayers, and in the intensity enhancement of shear modes due to the optical resonance with several optically allowed electronic transitions between conduction and valence bands in the band structures. This study paves way for fundamental understanding into the interface coupling of two-dimensional hybrids and heterostructures.}\end{abstract}

%\pacs{68.65.Pq 61.48.Gh 73.22.Pr 78.67.Wj}

\maketitle

Monolayer graphene (1LG) has a high carrier mobility and optical transparency, in addition to being flexible, robust and environmentally stable\cite{Geim,Bonaccorso}. Twisted bilayer graphene (t2LG) can be formed by stacking two 1LGs that are rotated with respect to each other\cite{Dos Santos,Kim,Reina}. In t2LGs, novel physical properties arise due to the periodically modulated interaction between the two Dirac electron gases with a large moir\'{e} supercell. Despite this modulation, a Dirac-like linear dispersion with a lower Fermi velocity than in 1LG is present\cite{Dos Santos,Ni,Trambly}. Moreover, the twist angle ($\theta_t$) between the layers can be used to tune the system properties such as optical absorption\cite{Moon}. The t2LG have parallel electronic bands across the Fermi level and the energy of the corresponding van Hove singularities (VHS) depends on $\theta_t$.\cite{Li} This is significantly different from the band structure of Bernal-stacked bilayer graphene (2LG) with a parabolic dispersion and parallel bands only within either valence or conductance bands.

In few-layer graphene (FLG), the band structure is distinct for each number of layers\cite{Koshino}. This can be exploited both for studying fundamental physics and device applications\cite{Zhu,Ye,Huang}.
By artificially assembling together $m$-layer graphene ($m$LG, $m\geq1$) and $n$-layer graphene ($n$LG, $n\geq1$), a ($m+n$)-layer graphene system [t$(m+n)$LG] can be formed. In this notation, the previously reported t2LG\cite{Dos Santos,Ni} can be denoted as t(1+1)LG. The $m$LG and $n$LG can be twisted by an angle $\theta_{t}$ with respect to each other. The tunability of t($m+n$)LG that arises from the number of layers and $\theta_t$ could lead to novel applications for these materials. The understanding of the fundamental physical properties, such as the strength of interlayer coupling between the two interface layers of $m$LG and $n$LG in $t(m+n)$LGs, and its impact on the electronic band structure and lattice dynamics, however, is crucial to achieve this goal.

As one of the most useful and versatile characterization tools, Raman spectroscopy has been widely used to probe the physical properties, such as phonons, electron-phonon interaction and band structure, in graphene and related materials\cite{Ferrari,Jorio,Ferrari2}. The first-order G band, arising from the emission of an in-plane optical phonon, can be easily observed around 1580~cm$^{-1}$ in 1LG, 2LG, FLG and bulk graphite\cite{Ferrari2}. The shear (C) mode present in FLGs\cite{Saha,Tan,Tsurumi}, and its low energy makes it sensitive to quasiparticles close to the Dirac point\cite{Tan}. However, it is usually not observed in typical experiments due to its weak intensity and low frequency (below 45~cm$^{-1}$). There are $n-1$ shear vibrational modes in $n$LG($n>$1). In contrast to other two dimensional materials such as MoS$_{2}$\cite{Zhang} and WSe$_{2}$\cite{Zhao}, only the highest-frequency shear mode has been observed in Bernal-stacked FLGs.\cite{Tan,Tsurumi} Due to the significantly weaker electron-phonon coupling, the detection of the other shear modes in Bernal-stacked FLG is challenging \cite{Tan}. Assembling 2D materials to form heterostructures and combined materials changes their symmetry, the lattice dynamics and electronic properties of the system. Raman spectroscopy of the shear modes provides a direct measurement of the interlayer coupling,\cite{Tan} offering information on the electronic and phonon properties of the system, as well as on their interplay. Thus, successfully probing the interlayer coupling in $t(m+n)$LGs via the C mode is a basic step to exploit physical properties of various combinations of two-dimensional materials.

Here, we study the C modes of t$(m+n)$LGs formed by twisting $m$LG and $n$LG with respect to each other to different angles. At a specific laser energy that depends on $\theta_t$, $m+n-2$ C modes arising from the $m-1$ and $n-1$ shear modes of the constituent FLGs are observed. The coupling between the graphene layers at the twisted interface is found to be about five times weaker than interlayer coupling in Bernal-stacked FLGs, and also the coupling to the layers adjacent to the interface is softened. The out-of-phase and in-phase vibrations of the two twisted, equivalent subsystems result in a Davydov splitting of $\sim$ 2~cm$^{-1}$ between bilayer C modes in t(2+2)LG. The C mode intensity enhancement in the t($m+n$)LG is shown to be due to an energy resonance between several VHSs of the joint density of states of all optically allowed transitions and laser excitations during the optical absorption and emission, confirmed by detailed theoretical calculations on the band structure of a t(1+3)LG. The VHSs contribute to different extent to the resonant profiles, which is a signal of varying electron-phonon coupling to different Raman bands.

\begin{figure}[tb]
\centerline{\includegraphics[width=90mm,clip]{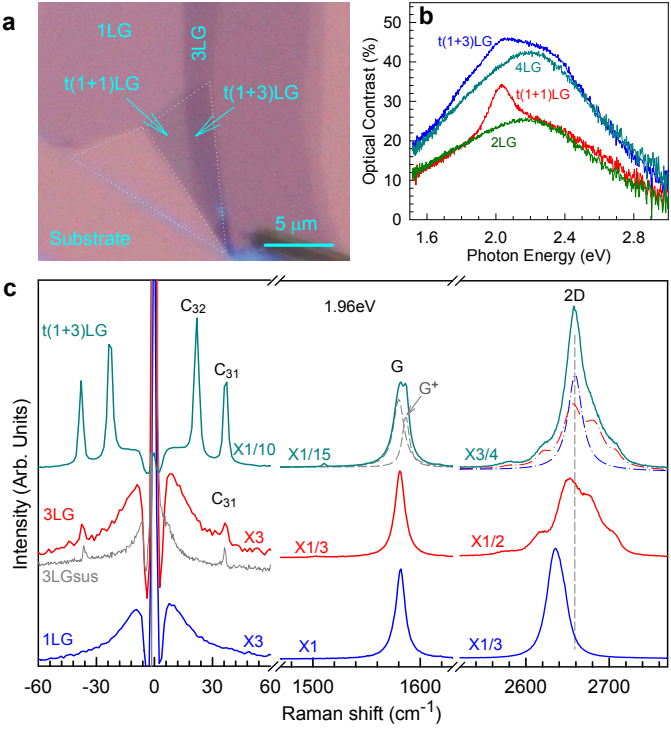}}
\caption{\textbf{Optical image, optical contrast and Raman spectra of t(1+3)LG.} (a) Optical image of a graphene flake containing t(1+1)LG and t(1+3)LG. (b) Optical contrast of t(1+1)LG and t(1+3)LG. For comparison, the contrast of 2LG and 4LG is also plotted. (c) Stokes/anti-Stokes Raman spectra of 1LG, 3LG and t(1+3)LG on SiO$_2$ in the C peak, and Stokes Raman spectra of 1LG, 3LG and t(1+3)LG in the G and 2D peak spectral regions at the 1.96eV excitation. The C mode of a suspended 3LG (3LG$_{\mathrm{sus}}$) is shown in gray for comparison. The t(1+3)LG 2D band is fitted by the 1LG- and 3LG-like 2D components as shown with dash-dotted lines, respectively. The G$^+$ mode is clearly identified by the fit to the t(1+3)LG G band.} \label{Fig1}
\end{figure}

\vspace*{5mm}
\noindent {\bf \large Results}

We have investigated Raman spectra of several t$(m+n)$LGs by various laser excitations. First, we focus on a typical graphene flake containing a 1LG, a trilayer graphene (3LG), a t(1+1)LG region and a t(1+3)LG region, the t(1+3)LG [t(1+1)LG] being formed by accidentally folding 1LG onto 3LG [1LG] during the mechanical exfoliation [see Fig.~\ref{Fig1}(a)]. The optical contrasts of t(1+1)LG and t(1+3)LG in Fig.~\ref{Fig1}(b) are, respectively, different from those of 2LG and 4LG. An additional feature appears at around 2.0eV for both t(1+1)LG and t(1+3)LG, revealing that the band structure of twisted few-layer graphene is modified after twisting in comparison to Bernal-stacked graphene layers.\cite{Havener,Wang}

Fig.~\ref{Fig1}(c) shows the Raman spectra of the flakes in the spectral ranges of the C, G and 2D bands, excited with the laser energy ($E_{\mathrm{ex}}$) 1.96~eV. All Raman spectra are normalized to the G intensity [I(G)] of 1LG. The 2D band of t(1+3)LG can be well fitted using the peak profiles of the 1LG and 3LG 2D bands, as indicated by the dashed-dotted lines in Fig.~\ref{Fig1}(c). The fitted 1LG-like and 3LG-like sub-peaks for t(1+3)LG are blueshifted by 24 and 5 cm$^{-1}$ relative to the ones in 1LG and 3LG, respectively. The blueshifts are attributed to the decrease of Fermi velocity in t(1+3)LG\cite{Trambly,Ni}, agreeing well with the previous results on t(1+1)LG\cite{Ni,Kim2}. This is a second indication that the 1LG and 3LG couple with each other after the formation of t(1+3)LG, modifying the band structure. The G band of t(1+3)LG is much stronger than that of 1LG and 3LG, suggesting a resonant enhancement of I(G) at the excitation energy ($E_{\mathrm{ex}}$) of 1.96~eV. A similar enhancement has been observed in t(1+1)LG \cite{Havener,Kim2}. However, in t(1+3)LG, the G peak exhibits a broader profile with a high frequency shoulder in comparison to the one reported for t(1+1)LG  \cite{Havener,Kim2} and in 3LG \cite{Tan}. We call this shoulder the $G^+$ peak.

In the low-energy region, the shear mode of the SiO$_2$-supported 3LG is observed at 37 cm$^{-1}$, with a worse signal-to-noise ratio than that of suspended 3LG (3LG$_{\mathrm{sus}}$) due to the strong background from the Si substrate.\cite{Tan} The t(1+3)LG contains four graphene layers. In Bernal-stacked 4LG, three C modes are expected at 41, 31 and 17~cm$^{-1}$.\cite{Tan} We use C$_{lk}$ to denote the $k^{\mathrm{th}}$ C mode of $l$LG, with $k$ increasing with a decreasing mode frequency. Thus, the 4LG C modes are denoted as C$_{41}$ (41~cm$^{-1}$), C$_{42}$ (31~cm$^{-1}$) and C$_{43}$ (17~cm$^{-1}$). However, only two ultralow-frequency modes at 22 cm$^{-1}$ and 37 cm$^{-1}$ are observed in t(1+3)LG, confirmed by the presence of their anti-Stokes components. Their frequencies agree with the frequencies of the two 3LG C modes predicted by a linear chain model\cite{Tan}. This suggests that the observed modes are related to C$_{32}$ and C$_{31}$ of the 3LG constituent in t(1+3)LG, respectively. C$_{32}$ and C$_{31}$ in t(1+3)LG are strongly enhanced with the intensity almost equal to I(G). The enhancement of the C and G peaks is a further indication of the coupling between 1LG and 3LG after twisting creates a heterostructure with properties, such as the band structure, that are distinct from its constituents.

\begin{figure*}[htb]
\centerline{\includegraphics[width=180mm,clip]{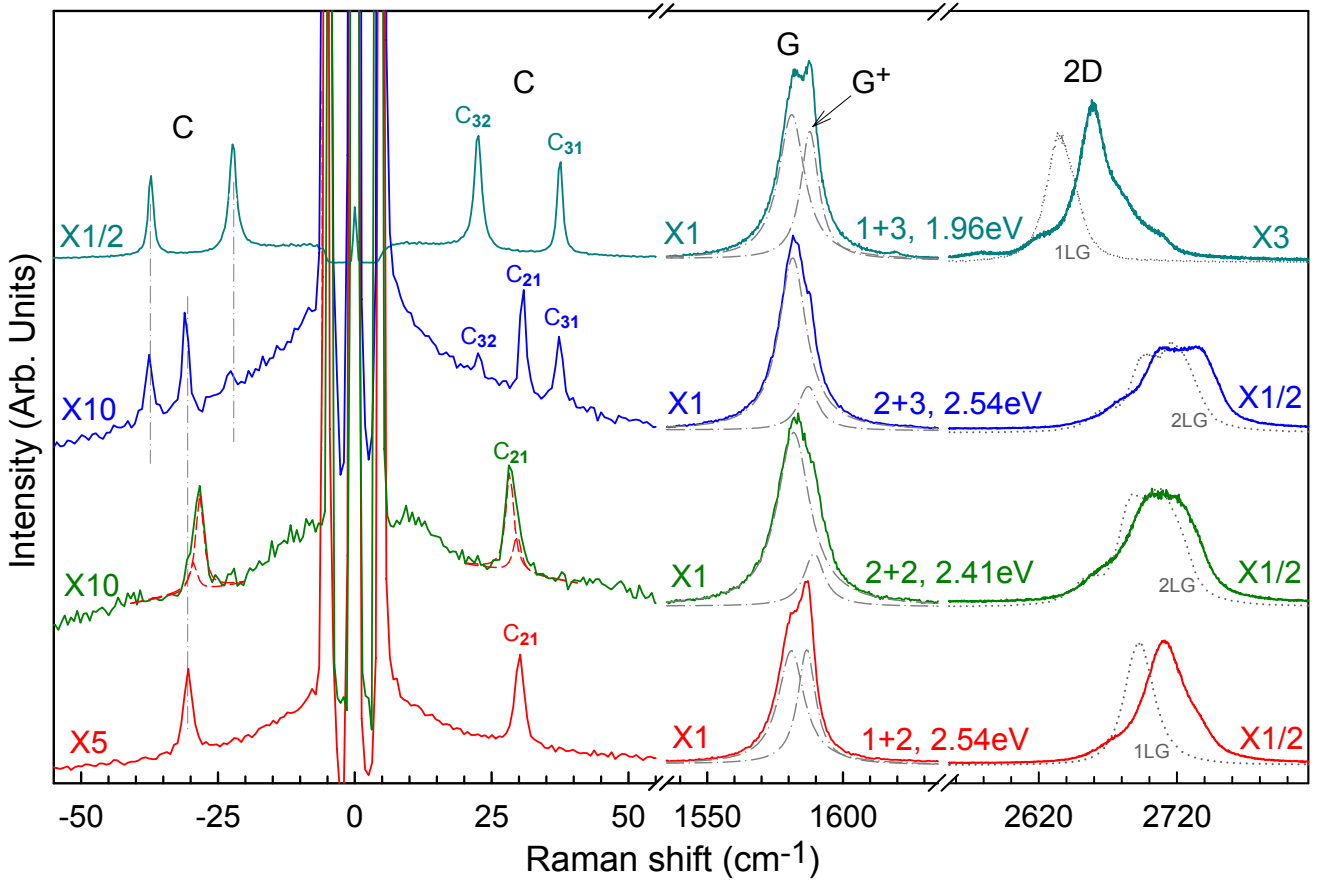}}
\caption{\textbf{Raman spectroscopy of t($m+n$)LG.} Stokes/anti-Stokes Raman spectra in the C peak region and Stokes spectra in the G and 2D spectral regions for four t($m+n$)LGs. For comparison, the 1LG and 2LG 2D peaks are plotted using dotted lines. The fits show that the G band of each t($m+n$)LG is composed of two subpeaks, G and G$^+$. The excitation energy $E_{\mathrm{ex}}$ used for each t($m+n$)LG is indicated. The spectra are scaled and offset for clarity. The scaling factors of the individual spectra are shown on the left.} \label{Fig2}
\end{figure*}

\begin{figure*}[htb]
\centerline{\includegraphics[width=180mm,clip]{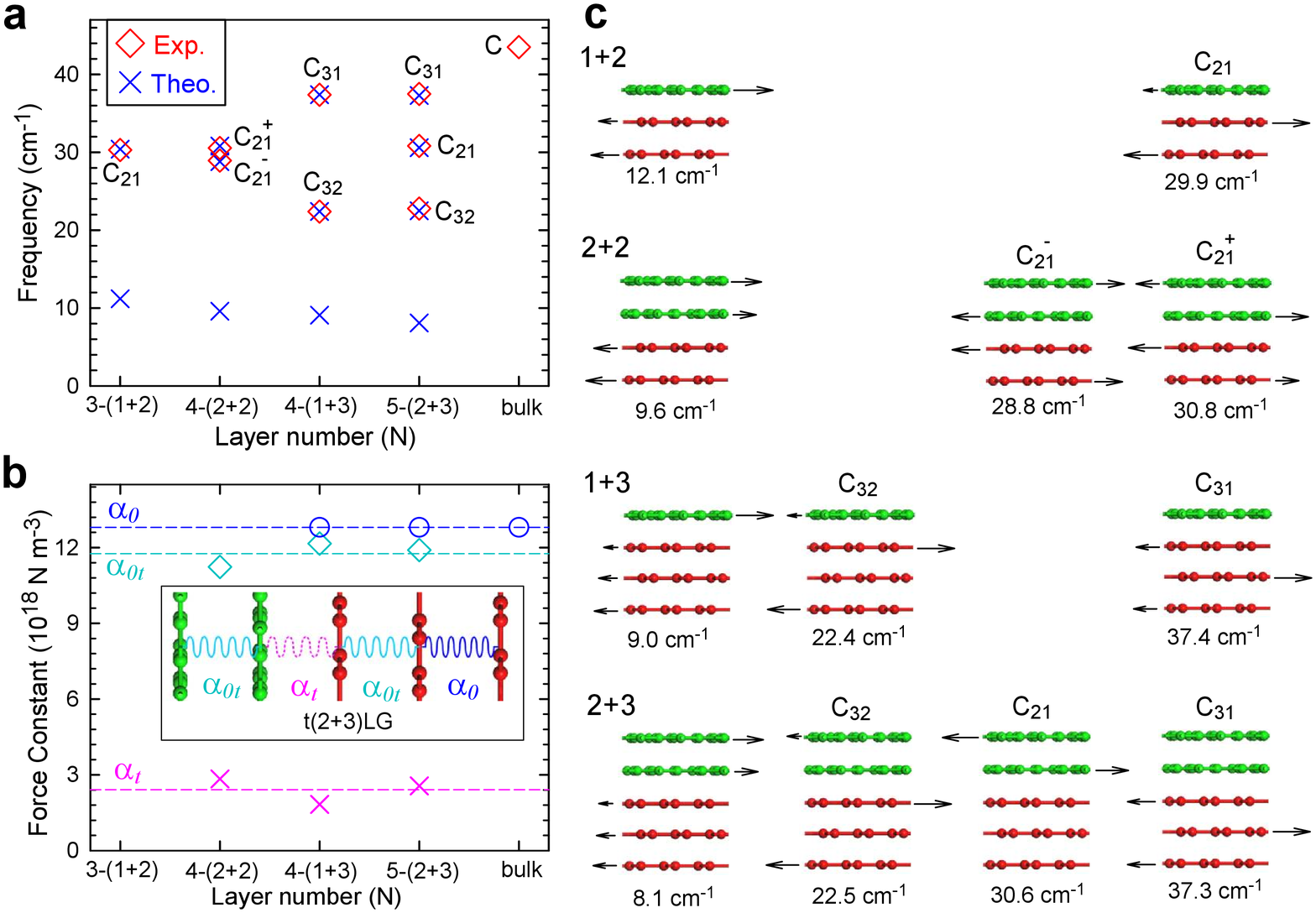}}
\caption{\textbf{Shear mode frequencies and normal mode displacements in t($m+n$)LG.} (a) Experimental (Exp., open diamonds) and theoretical (Theo., crosses) frequencies of the C modes in t($n+m$)LGs. (b) Fitted shear force constants ($\alpha$) in t($n+m$)LGs. The inset shows a schematic diagram of an improved linear chain model for the t(2+3)LG with the bulk interlayer force constant $\alpha_0$, interface force constant $\alpha_t$, and the softened force constant adjacent to the interface $\alpha_{0t}$. (c) Normal mode displacements and mode frequencies of the shear modes calculated using the improved linear chain model.} \label{Fig3}
\end{figure*}

The observation of C$_{31}$ and C$_{32}$ in t(1+3)LG also shows that the weaker coupling between 1LG and 3LG in t(1+3)LG makes its lattice dynamics different from the Bernal-stacked 4LG. This is also true for other t($m+n$)LGs. Fig.~\ref{Fig2} shows the Stokes (S) and anti-Stokes (AS) spectra at the C region and Stokes spectra at G and 2D spectral regions for four t($m+n$)LGs with $m$=1 or 2 and $n$=2 or 3 (See Supplementary Fig.1 for optical images). To facilitate comparison, all spectra are normalized to their respective I(G). The 2D bands of all t($m+n$)LGs blue-shift relative to the 1LG or 2LG ones measured under the same conditions due to the coupling, similar to the case of t(1+1)LG reported in the literature\cite{Ni}.  Only one mode located at $\sim$30~cm$^{-1}$ is observed for the t(1+2)LG with its frequency close to the C mode of the 2LG constituent, C$_{21}$. In contrast, t(2+2)LG exhibits a broad and asymmetrical peak. It can be fitted using two Lorentzian peaks at 28.9 cm$^{-1}$ and 30.6 cm$^{-1}$, denoted as C$_{21}^-$ and C$_{21}^+$, respectively. In t(2+3)LG, besides the two C modes of its 3LG constituent, an additional mode at 30.8 cm$^{-1}$ related to its 2LG constituent is present. In t(1+3)LG, as shown in Fig.~\ref{Fig1} and discussed above, two C modes related to its 3LG constituent are observed. In short, $m+n-2$ C modes are observed in a t($m+n$)LG for specific excitation energies, instead of $m+n-1$ C modes expected in Bernal-stacked ($m+n$)LGs. The observed modes are related to the C modes of the constituent $m$LG (if $m>$1) and $n$LG ($n>$1). The frequencies of all C modes shown in Fig.~\ref{Fig2} are summarized in Fig.~\ref{Fig3}(a) using open diamonds.

Similar to t(1+3)LG, a double peak structure in the G band with an additional G$^{+}$ peak is observed for all t($m+n$)LGs, as revealed by the Lorentzian fits shown in Fig.~\ref{Fig2}. The G$^+$ peak is observed neither in Bernal-stacked 1LG, 3LG, 4LG nor t(1+1)G. Thus, the G$^+$ mode in t($m+n$)LG must arise from the multilayered constituents and their interaction. The G peak position is expected to be sensitive to the layer number of FLG\cite{Saha,Wang2}, which is in contrast to the experimental result\cite{Tan}. Here, the observed G$^+$ position in t($m+n$)LGs is 6-7 cm$^{-1}$ higher than the corresponding G peak, also independent on the layer number. This value is identical to that between Raman-active E$^2_{2g}$ and infrared-active E$_{1u}$ modes in bulk graphite. Therefore, we prefer to attribute the G$^+$ peak in t($m+n$)LGs to the zone-center infrared-active E$_{u}$ or $E''$ modes in their 2LG and 3LG constituents.

To understand the interlayer coupling in t($m+n$)LG, an improved linear chain model is introduced. The frequencies $\omega$ (in cm$^{-1}$) and displacement patterns of the shear modes in t($m+n$)LG can be calculated by solving linear homogeneous equations as follows,\cite{Born}
\begin{equation}
\omega_{i}^2\mathbf{u}_{i}=\frac{1}{2{\pi }^{2}{c}^{2}\mu }\mathbf{D}\mathbf{u}_{i},
\label{eq:displ}
\end{equation}
\noindent where $\mathbf{u}_{i}$ is the phonon eigenvector of the mode $i$ with a frequency $\omega_i$, $\mu$=7.6$\times$10$^{-27}$~kg$\AA^{-2}$ the monolayer mass per unit area, $c$ the speed of light in cm s$^{-1}$, and $\mathbf{D}$ the shear part of the force constant matrix (for details, see Supplementary Note 1). Here, only nearest-neighbor interlayer interaction is assumed.\cite{Tan} The force constant per unit area between bulk layers away from the interface is given by $\alpha_0$. In FLGs, this interlayer force constant was experimentally determined to be $\alpha_0$=12.8$\times$10$^{18}$ Nm$^{-3}$.\cite{Tan} We denote the interlayer shear force constant between two twisted interface layers ($m$LG and $n$LG) by $\alpha_t$. We also assume that the presence of the interface perturbs the force constant between the two layers closest to the interface in both subsystems, described by the force constant $\alpha_{0t}$. The force constants are illustrated in the inset of Fig.~\ref{Fig3}(b) for a t(2+3)LG case.

Based on the experimental C mode frequencies of t(2+2)LG, t(1+3)LG and t(2+3)LG, we fit the parameters ($\alpha_{0t}$ and $\alpha_{t}$) of the improved linear chain model. The average values for $\alpha_{0t}$ and $\alpha_{t}$ are, 11.8$\times$10$^{18}$~Nm$^{-3}$ and 2.4$\times$10$^{18}$~Nm$^{-3}$, respectively. These values suggest that the interlayer coupling between the two twisted interface layers is approximately 5 times weaker than the coupling in the Bernal-stacked layers, while the coupling between layers next to the interface, $\alpha_{0t}$, decreases by approximately 9~\% with respect to $\alpha_0$. To validate the fitting procedure, we calculate the position of C$_{21}$ of t(1+2)LG using these parameters. The calculated value of 30.4~cm$^{-1}$ is in good agreement with the experiment.

The theoretical and experimental C mode frequencies in t($m+n$)LGs are summarized in Fig.~\ref{Fig3}(a) as crosses and open diamonds, respectively. Fig. \ref{Fig3}(c) shows the normal mode displacements of the C modes in t($m+n$)LGs calculated using Eq.~(\ref{eq:displ}). The mode displacements for the C modes are mainly localized within the $n$LG or $m$LG constituents and they are only weakly affected by the coupling across the twisted interface. Due to the interlayer coupling between the layers at the twisted interface, also a low-frequency mode below 15 cm$^{-1}$ is theoretically predicted. This mode corresponds to the relative vibration of $m$LG and $n$LG constituents in t($m+n$)LGs, as illustrated in Fig.~\ref{Fig3}(c). It is, however, not experimentally observed in the present measurements, possibly due to its weak intensity caused by a weak electron-phonon coupling.

In Fig.~\ref{Fig2}, two subpeaks, split by $\approx$ 2~cm$^{-1}$, are found for the t(2+2)LG shear mode. It is interesting to compare the lattice dynamics of t(2+2)LG and Bernal stacked 4LG. The C$_{41}$ ($E_g$) mode of 4LG at 41~cm$^{-1}$ does not exist in t(2+2)LG because of the weak coupling at the interface. The C$_{42}$ ($E_u$) mode of 4LG is infrared-active. For this mode, two middle layers move in phase, and the two bilayer systems are out-of-phase with respect to each other. This effectively makes the system equivalent to two uncoupled bilayers. Thus, the C$_{42}$ frequency (31cm$^{-1}$) of 4LG is equal to that of the C mode in 2LG.\cite{Tan} For t(2+2)LG, the displacements of the middle layers are in-phase for C$_{21}^-$. Its lower frequency (28.8cm$^{-1}$) than 2LG C$_{21}$ and 4LG C$_{42}$ frequency directly suggests a softening of the coupling between layers next to the interface. C$_{21}^+$ has a higher frequency than C$_{21}^-$ because the weakly coupled middle layers vibrate out-of-phase\cite{Wieting}. The frequency difference between C$_{21}^-$ and C$_{21}^+$ is referred to as Davydov splitting.\cite{Ghosh} Such splitting is a direct signature of the presence of coupling, albeit weak, between the two interface layers. Actually, the weak interlayer coupling at the interface can be confirmed by the frequencies of three C modes in t(2+2)LG. Indeed, the frequency of the relative vibration between the top and bottom bilayers in t(2+2)LG can be approximately deduced\cite{Wieting} as $\sqrt{\omega^2_{C_{21}^+}-\omega^2_{C_{21}^-}}$ = 10.9 cm$^{-1}$, which agrees well with the predicted 9.6 cm$^{-1}$. A similar Davydov splitting of the C mode is expected to occur in all t$(n+n)$LG.

\begin{figure}[htb]
\centerline{\includegraphics[width=90mm,clip]{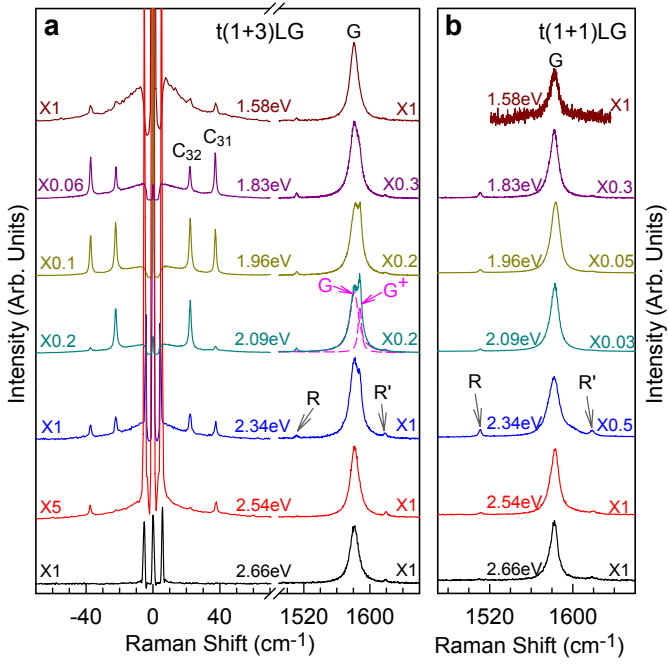}}
\caption{\textbf{Raman spectroscopy of t(1+3)LG and t(1+1)LG at different excitation energies ($E_{\mathrm{ex}}$).}(a) The C and G modes of t(1+3)LG and (b) the G mode of t(1+1)LG at seven different $E_{\mathrm{ex}}$. The R and R$'$ modes are indicated by arrows. The presence of the $G^+$ mode is revealed by the two-Lorentzian fit to the t(1+3)LG G band at $E_{\mathrm{ex}}$=2.09eV. The spectra are scaled and offset for clarity. The scaling factors of the individual spectra are shown on the left.}\label{Fig4}
\end{figure}

The C and G$^+$ modes of each t($m+n$)LG can only be detected in a small $E_{\mathrm{ex}}$ range (see the Supplementary Fig.2), suggesting that the modes are enhanced in intensity by specific values of $E_{\mathrm{ex}}$. Here, we consider in detail the case of t(1+3)LG. 13 laser wavelengths were used to excite the C and G modes of t(1+3)LG. As a simple model system for t($m+n$)LGs, the Raman spectra of t(1+1)LG are included at multiple excitation wavelengths. Fig.~\ref{Fig4} shows the measured spectra at seven different wavelengths both for t(1+1)LG and t(1+3)LG, with peak intensities normalized to I(G) in 1LG. The intensity of the G mode in t(1+1)LG and that of the C, G and G$^+$ modes in t(1+3)LG strongly depend on $E_{\mathrm{ex}}$, showing a typical resonant behavior. Fig.~\ref{Fig4}(a) clearly shows that the G$^+$ peak in t(1+3)LG can only observed in a small range of $E_{\mathrm{ex}}$.

In order to explain the physical origin of the resonant enhancement, to be extended to all t($m+n$)LG\cite{Tan2}, a detailed calculation of the band structure in t(1+1)LG and t(1+3)LG is necessary. As the twist angle $\theta_t$\cite{Carozo} determines the coupling between the constituent $n$LG and $m$LG and consequently the intensity enhancement, its accurate determination is crucial. The twist angle can also be expressed in terms of the twist vector\cite{Trambly,Sato} $(p,q)$. In the literature, no Raman-based methods to determine the twist angle in arbitrary t($n+m$)LG have been presented. However, the so-called R and R' Raman bands have been used to determine the angle in t(1+1)LG\cite{Carozo,Carozo2}. As shown in Fig.~\ref{Fig1}(a), in the present experiment, a graphene layer is flipped over and folded onto 1LG and 3LG regions to form the t(1+1)LG and t(1+3)LG. Consequently, as the top and bottom layers in t(1+1)LG and t(1+3)LG are actually the same graphene sheet, both systems have the same $\theta_t$. The R and R$'$ peaks of t(1+1)LG are, respectively, found at 1510 cm$^{-1}$ and 1618~cm$^{-1}$. A further confirmation for the same twist angle is the observation of the t(1+3)LG R and R$'$ bands at 1512 cm$^{-1}$ and 1618~cm$^{-1}$. From the position of the R band\cite{Carozo,Kim2, Carozo2}, $\theta_t$ can be determined as $\sim 10.6^{\circ}$, that is close to a commensurate structure with a twist vector of (1,9) and $\theta_t = 10.4^{\circ}$. Indeed, the optical contrast of t(1+1)LG in Fig. ~\ref{Fig1}(b) shows an absorption peak in 2.03 eV, which agrees with the calculated one \cite{Carozo2} for (1,9) t(1+1)LG. \cite{Carozo2,Trambly} We thus use this commensurate structure in calculations both for t(1+1)LG and t(1+3)LG.

\begin{figure*}[htb]
\centerline{\includegraphics[width=180mm,clip]{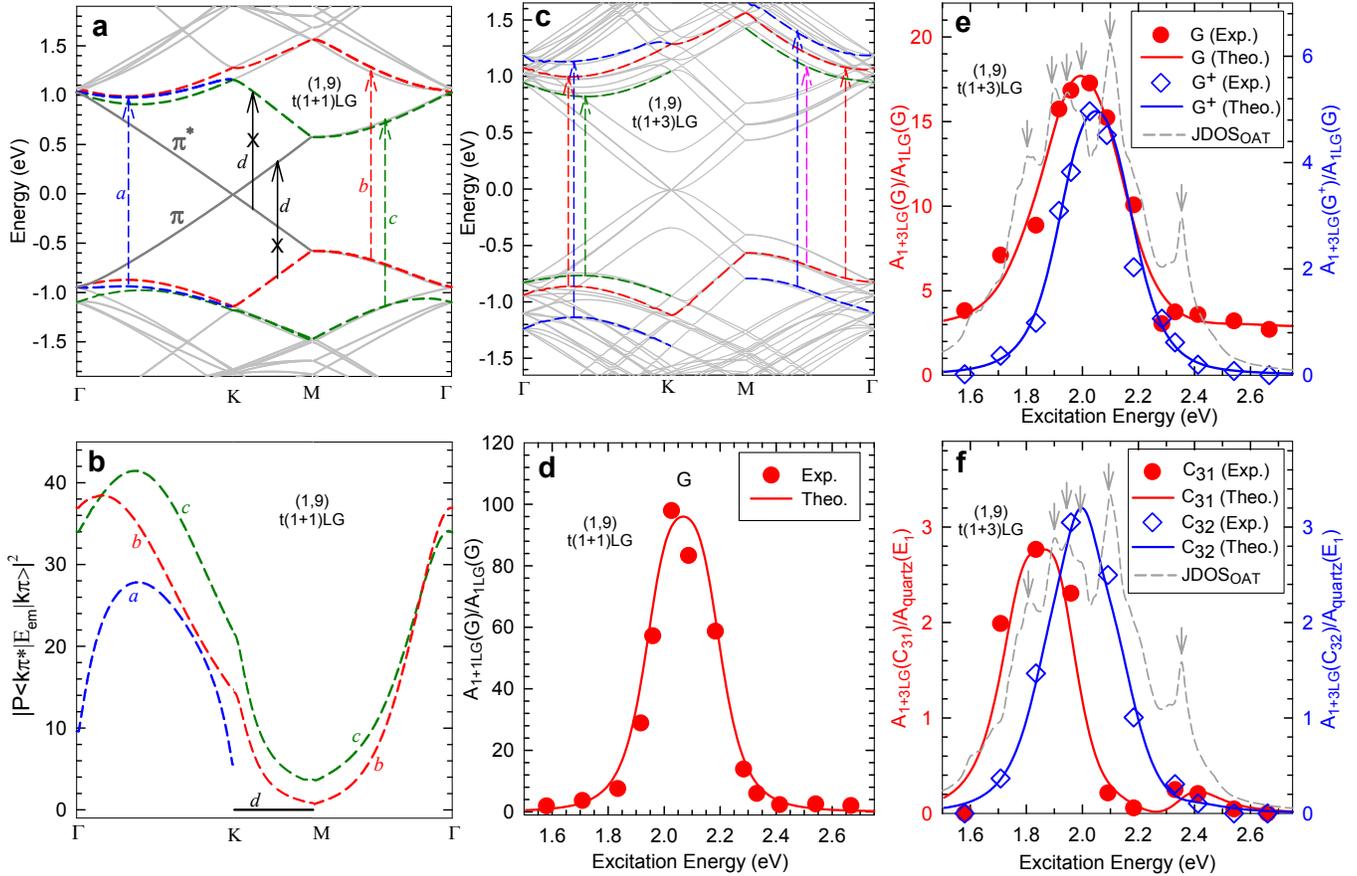}}
\caption{\textbf{Intensity of the C and G modes.} (a) The band structure of (1,9) t(1+1)LG. The optically allowed transitions are marked by dashed arrows. The transitions with the energy of $\sim$ 1.15eV between parallel bands along K-M direction are forbidden, as indicated by the solid arrows with crosses. (b) Squared optical matrix elements of the corresponding band pairs in (a). (c) The band structure of (1,9) t(1+3)LG. Some typical transitions are indicated by vertical dashed lines. (d) A(G) of t(1+1)LG, (e) A(G) and A(G$^{+}$) of t(1+3)LG, and (f) A(C$_{31}$) and A(C$_{32}$) as a function of $E_{\mathrm{ex}}$. The filled circles and open diamonds show the experimental data, and the solid lines the simulations. The gray dash-dotted lines in (e) and (f) are the JDOS$_{OAT}$ in t(1+3)LG along $\Gamma$-K-M-$\Gamma$. The VHSs are indicated with arrows.} \label{Fig5}
\end{figure*}

The band structure of (1,9) t(1+1)LG calculated using the DFTB+ program \cite{Aradi,Zhang2} is shown in Fig. ~\ref{Fig5}(a). Fig.~\ref{Fig5}(b) shows the squared optical matrix elements for electronic transitions of some pairs between conduction and valence bands in Fig. ~\ref{Fig5}(a) (see Methods for computational details). The optically allowed transitions are indicated by dashed lines with arrows. Obviously, the optical matrix elements are anisotropic along the high-symmetry directions as illustrated in Fig.~\ref{Fig5}(b). In particular, the optical transitions of (1,9) t(1+1)LG between two parallel conduction and valence bands labeled by the two arrows with crosses are found to be forbidden, and thus they do not contribute to the intensity resonance of Raman modes.

The band structure of (1,9) t(1+3)LG is shown in Fig.~\ref{Fig5}(c). In the low-energy region, two parabolic bands from the 3LG still exist in the band structure of t(1+3)LG. However, a linear band arising from the 1LG constituent is superimposed with that of the 3LG constituent to form a doubly degenerate linear band in t(1+3)LG, with a Fermi velocity that is decreased by $\sim$4\% compared to that in 1LG due to the coupling at the interface. This reduction manifests itself in a blueshift in frequency of the 1LG-like 2D peak as shown in Fig.~\ref{Fig1}(c). The selection rule for optical transitions in t(1+3)LG is more complex than that in t(1+1)LG. As an example, some typical optically allowed transitions in t(1+3)LG are shown in Fig.~\ref{Fig5}(c) with vertical arrows. This results in a broader absorption peak in t(1+3)LG in comparison to the case in t(1+1)LG as shown in Fig.~\ref{Fig1}(b).

The experimental areal intensities of the G band in t(1+1)LG [A(G)] and the G and C bands in t(1+3)LG [A(G),A(G$^+$), A(C$_{31}$) and A(C$_{32}$)] as a function of $E_{\mathrm{ex}}$ are, shown in Figs.~\ref{Fig5}(d), ~\ref{Fig5}(e) and ~\ref{Fig5}(f), respectively. The G modes were normalized to A(G) of 1LG, whereas the shear modes were normalized to the quartz $E_{1}$ modes (at 127 cm$^{-1}$)\cite{Krishnan} to eliminate the effect of different CCD efficiency for the detection of C and G modes at each $E_{\mathrm{ex}}$. To understand the experimentally observed intensity enhancement, the Raman intensity of the C and G modes was calculated by second order perturbation theor\cite{Saito}. Because only the optically allowed electronic transitions can be involved in the resonant Raman process, we calculate the electronic joint density of states (JDOS) of all the optically allowed transitions (JDOS$_{OAT}$) in t($m+n$)LG by the following equation:
\begin{equation}
JDOS_{OAT}\left({E}\right)\propto\sum_{ij}\sum_{k}{\left|M_{ij}(\mathbf{k})\right|}^{2}\delta\left({E}_{ij}\left(\mathbf{k} \right)-{E}\right),
\label{eq:jdos}
\end{equation}
\noindent where $M_{ij}(\mathbf{k})$ is the optical matrix element between
the $i^{\mathrm{th}}$ conduction and $j^{\mathrm{th}}$ valence bands, $E_{ij}(\mathbf{k})$ gives the transition energy of $i \rightarrow j$ band pair at wavevector $\mathbf{k}$. 180 $\mathbf{k}$-points are used for the summation along the $\Gamma$-K-M-$\Gamma$ path, distributed according to the respective path lengths in the reciprocal space. The optically allowed transitions in t(1+1)LG shown in Fig.~\ref{Fig5}(a) contribute to a VHS at 1.95eV with the full width at half maximum (FWHM) of 0.13 eV in the JDOS (see Supplementary Fig.3). The JDOS$_{OAT}$ in t(1+3)LG is shown in Fig.~\ref{Fig5}(e,f) by gray dashed lines. There are six distinctive VHS features in the JDOS$_{OAT}$ of (1,9) t(1+3)LG, which are labeled by gray arrows(see the Supplementary Fig.4). t(1+3)LG exhibits much broader JDOS$_{OAT}$ (FWHM$\approx$0.5 eV) than t(1+1)LG because of its complex band structure. The major contribution for the intensity comes from resonance matching between $E_{\mathrm{ex}}$ and the energy of VHSs in the JDOS$_{OAT}$ of t($m+n$)LG. Therefore, the Raman intensity of Raman modes in t($m+n$)LG as a function of $E_{\mathrm{ex}}$ can be evaluated as\cite{Carozo2},
\begin{equation}
I\propto{\left|\sum_{j}\frac{{M}_{j}}{\left(E_{\mathrm{ex}}-{E_{\mathrm{VHS}}}(j)-i\gamma\right)\left(E_{\mathrm{ex}}-{E}_{ph}-{E_{\mathrm{VHS}}}(j)-i\gamma\right)} \right|}^{2},\\
\label{eq:inten}
\end{equation}
\noindent where $M_j$ are constants treated as fitting parameters that encompass the product of the electron-phonon and electron-phonon interaction matrix elements for the $j^{th}$ VHS in the JDOS$_{OAT}$, $E_{ph}$ is the phonon energy (0.196 eV for $G$, 0.197 eV for $G^{+}$, 4.6 meV for $C_{31}$, 2.7 meV for $C_{32}$), and $\gamma$ gives the energy uncertainty related to the lifetime of the excited state (here, $\gamma$=0.15 eV ).  The calculated intensity of the C and G bands as a function of $E_{\mathrm{ex}}$ are shown in Fig.~\ref{Fig5}(d)-\ref{Fig5}(f) with solid curves. Both well agree with the experimental measurements.

The resonant profile of the G$^+$ peak of t(1+3)LG is symmetric and centered at 2.05 eV with a width of 0.3eV, which indicates that the VHSs with lower ($<$1.8eV) and higher ($>$2.3eV) energies contribute little to the resonance of the G$^+$ peak. However, that of the G peak is asymmetric and slightly broader, meaning that VHSs with wider energy range contribute to its resonance, particularly in the lower energy range ($<$1.9eV). When out of resonance, the G$^+$ peak vanishes while I(G) of t(1+3)LG is about 3 times as much as I(G) of 1LG.

Because of the small energy of the C mode phonon, the resonance condition for both incoming and outgoing photon can be simultaneously nearly fulfilled. This makes the C mode intensity of t(1+3)LG be significantly enhanced by the optically allowed transitions in t(1+3)LG, as demonstrated in  Figs.~\ref{Fig5}(c) and ~\ref{Fig5}(f). Indeed, I(C$_{31}$) and I(C$_{32}$) excited at 1.96 eV can be enhanced by about 120- and 190-fold compared to I(C$_{31}$) of Bernal-stacked 3LG, respectively, and their intensities can be comparable to resonantly enhanced I(G) and I(G$^+$). In particular, both $C_{31}$ and $C_{32}$ exhibit broad resonant profiles with a width comparable to the G and G$^+$ peaks. The resonant profile of $C_{32}$ is blue-shifted by 0.13 eV relative to that of $C_{31}$. I($C_{32}$) is resonantly enhanced by the transitions associated with VHSs at 2.1 eV, which contribute little to I($C_{31}$), while I($C_{31}$) is significantly enhanced by the transitions associated with VHSs below 1.8 eV.

For an electronic transition at some wavevector $\mathbf{k}$, the matrix elements related with electron-photon interactions are the same for C$_{31}$ and C$_{32}$. The EPC matrix element, however, can be different. Indeed, the strength of the EPC greatly varies for the different graphene phonon branches.\cite{Lazzeri} As the EPC of C$_{41}$ in 4LG is about 15 times larger than the other Raman-active C$_{43}$ mode at 17 cm$^{-1}$, the other C modes except C$_{n1}$ in $n$LG are challenging to detect.\cite{Tan} The different resonant behaviors between C$_{31}$ and C$_{32}$ of 3LG had been observed in scroll structures at edges.\cite{Tan2} Thus, the observation of different resonant profiles for C$_{31}$ and C$_{32}$ directly indicates that in addition to differences in the EPC strength, the EPC is also anisotropic in the $\mathbf{k}$ space, thus effectively removing the contribution from some optically allowed transitions to the resonant Raman process of some vibration modes.

\vspace*{5mm}
\noindent {\bf \large Discussion}

The C mode of Bernal-stacked FLG can be well fitted with a Breit-Wagner-Fano (BWF) lineshape.\cite{Tan} This is attributed to quantum interference between the C mode and the continuum of electronic transitions near the K point.\cite{Tan} Our theoretical calculation on the band structure of t(1+3)LG reveals that twisting can not induce a band gap near the K point. Therefore, a similar BWF profile is present for the C mode in the twisted system and its intensity is of the same order of magnitude as that of the C mode in Bernal-stacked 3LG. However, the intensity enhancement of the C mode in t(1+3)LG masks the weak BWF profile resulting from quantum interference between the C phonon and the continuum of electronic transitions at K point, giving an overall Lorentzian profile.

The weak coupling between the two interface layers in t($m+n$)LGs can be understood from the nature of C modes that originate from the interlayer shear vibration. Due to the twist, the atoms at the interface are misaligned with respect to each other, which results in a weaker moir{\'{e}}-modulated coupling between the two interface layers, compared to the high order in conventional AA, AB or ABC stacked FLG with carbon atoms on the top or hollow positions of the neighboring layer.

Because the frequencies of the shear mode phonons directly reflect the interlayer interactions, the measurement of the C modes in t($m+n$)LG allows the direct evaluation of interlayer interactions, both at the twisted interface and in the neighboring layers. The coupling across this interface is much weaker than the interaction between Bernal-stacked layers. Nevertheless, such a weak interaction can modify the electronic dispersion, as proven by the resonant enhancement of the C modes. By measuring the C modes in two-dimensional hybrids and heterostructures, the interlayer interactions at their interface can be probed in a similar manner.

Note added. Recently, we became aware of a reprint reporting experimental works on the intensity enhancement of the C modes in t($n+n$)LG.\cite{Yu-2014}

\vspace*{5mm}
\noindent {\bf \large Methods}

\noindent {\bf Sample preparation.} Highly oriented pyrolytic graphite (HOPG) is mechanically exfoliated on a Si substrate covered with a 90nm SiO$_2$ layer to obtain multilayer graphene \cite{Novoselov}. To form t($m+n$)LG, a $m$LG flake is accidentally flipped over and folded onto a $n$LG flake  during the exfoliation process, or a $m$LG flake from one Si substrate is transferred onto a $n$LG flake on another Si substrate.\cite{Reina2} The number of layers in all initial and twisted graphene flakes is identified by Raman spectroscopy and optical contrast.\cite{Casiraghi,Zhao2}

\noindent {\bf Raman measurements.} Raman spectra are measured in a back-scattering geometry at room temperature with a Jobin-Yvon HR800 Raman system, equipped with liquid-nitrogen-cooled charge-coupled device, a 100$\times$ objective lens (NA=0.90) and several gratings. The excitation energies used are 1.58eV and 1.71eV of a Ti:Saphire laser, 1.96eV, 2.03eV, 2.09eV and 2.28eV of a He-Ne laser, and 1.83eV, 1.92eV, 2.18eV, 2.34eV and 2.41eV of an Kr$^+$ laser, and 2.54eV and 2.67eV of an Ar$^+$ laser. The resolution of the Raman system at the 2.41 eV excitation is 0.54 cm-1 per CCD pixel. The laser excitations are cleaned away from their plasma lines with BragGrate Bandpass filters. Measurements of the Raman shift with frequencies close to 5 cm$^{-1}$ for each excitation are enabled by three BragGrate notch filters with optical density 3 and with FWHM of 5-10 cm$^{-1}$.\cite{Tan} Both BragGrate bandpass and notch filters are produced by OptiGrate Corp. The typical laser power was $\sim$0.5~mW to avoid sample heating.

\noindent {\bf Band structure calculations.} The electronic structure calculations were carried out using the DFTB+ program\cite{Aradi,Zhang2}. DFTB+ is an implementation of the Density Functional based Tight Binding (DFTB) method, containing many extensions to the original method. DFTB is based on a second-order expansion of the Kohn-Sham total energy in Density-Functional Theory (DFT) with respect to charge density fluctuations. The Coulomb interaction between partial atomic charges is determined using the self-consistent charge (SCC) formalism. A Slater Kirkwood-type dispersion is employed for the van der Waals and $\pi$-$\pi$ stacking interactions. Here we adopted the parameter set 'mio-0-1'\cite{Elstner} for Slater-Koster files. This approach has been shown to give a reasonably good prediction of the band structure in graphene and its derivatives\cite{Zheng,Gao}.

\noindent {\bf Optical matrix elements.} To calculate the optical matrix elements, the tight-binding model proposed by Trambly de Laissardi{\`{e}}re \textit{et al.}\cite{Trambly2, Trambly} and recently used for the calculation of optical properties in twisted bilayer graphene\cite{Moon} was used to calculate the electronic band structure of t($n+m$)LG. The model hopping parameters were scaled up by 18~\% to compensate for the underestimation of Fermi velocity in the absence of corrections due to electron-electron interaction.\cite{Coh}

The optical matrix elements were calculated as\cite{Venezuela}
\begin{equation}
\langle \mathbf{k} \pi_i^{\star()} | E_{\mathrm{e-m}}^{\mathrm{in [out]}}|  \mathbf{k} \pi_j^{(\star)} \rangle = \mathbf{P}_{\mathrm{in [out]}} \cdot \langle \mathbf{k} \pi_i^{\star()} | \nabla H_{\mathbf{k}}|  \mathbf{k} \pi_j^{(\star)} \rangle,
\label{eq:opt_el}
\end{equation}
\noindent where $\pi_i^{\star()}$ denotes the $i^{\mathrm{th}}$ state in the valence (conduction) band and $\mathbf{P}_{\mathrm{in [out]}}$ is the light polarization. $\mathbf{P} = [1 1]$ in the present calculations.

\vspace*{5mm}
\noindent {\bf \large Acknowledgements}

\noindent We acknowledge support from the National Natural Science Foundation of China, grants 11225421, 10934007, 11434010 and 11474277. ACF acknowledges funding from EU projects GENIUS, CARERAMM, MEM4WIN, EU Graphene Flagship (contract no. 604391), ERC grant Hetero2D, a Royal Society Wolfson Research Merit Award, EPSRC grants EP/K01711X/1, EP/K017144/1, EP/L016087/1.

\vspace*{5mm}
\noindent {\bf \large References}
%\bibliographystyle{naturemag}
%\bibliography{C-twisted}

\end{document}